\documentclass[fleqn,12pt,twoside]{article}
\usepackage[headings]{espcrc1}
\usepackage{epsfig}

\readRCS
$Id: qm05ps_v1.8.tex,v 1.1 2005/10/10 14:25:59 ufstasze Exp $
\usepackage{graphicx}
\usepackage[figuresright]{rotating}

\newcommand{\rootsnn}[1]{$\sqrt{s_{NN}} = #1$~GeV}
\newcommand{\ene}[1]{$#1$~GeV}
\newcommand{\pt}{$p_{T}$ }
\newcommand{\ptn}{$p_{T}$}
\newcommand{\dndeta}{dN$_{ch}$/d$\eta$ }
\newcommand{\pionr}{$N_{\pi^{-}}/N_{\pi^{+}}$ }
\newcommand{\pir}{$N_{\pi^{-}}/N_{\pi^{+}}$}
\newcommand{\kaonr}{$N_{K^{-}}/N_{K^{+}}$ }
\newcommand{\kar}{$N_{K^{-}}/N_{K^{+}}$}
\newcommand{\protr}{$N_{\bar{\rm p}}/N_{\rm p}$ }
\newcommand{\pr}{$N_{\bar{\rm p}}/N_{\rm p}$}
\newcommand{\Raa}{$R_{AA}$ }
\newcommand{\Rnn}{$R_{AA}$}

\newcommand{\Rcp}{$R_{CP}$ }
\newcommand{\Rncp}{$R_{CP}$}
\newcommand{\Npart}{N$_{part}$ }
\newcommand{\Ncoll}{N$_{coll}$ }
\newcommand{\Nnpart}{N$_{part}$}
\newcommand{\Nncoll}{N$_{coll}$}

\newcommand{\AmS}{{\protect\the\textfont2
  A\kern-.1667em\lower.5ex\hbox{M}\kern-.125emS}}

\title{Recent Results from the BRAHMS Experiment}

\author{P. Staszel\address[KRA]{Jagiellonian University, Institute of Physics, \\
        ul. Reymonta 4, 30-059 Krak{\'o}w, Poland}
        (for the BRAHMS\thanks{For the full BRAHMS Collaboration 
         author list and acknowledgment see appendix 'Collaborations' of this volume} Collaboration)
       }
       
\runtitle{Recent Results from the BRAHMS Experiment}
\runauthor{P. Staszel}

\begin{document}

\maketitle

\begin{abstract}

We present recent results obtained by the BRAHMS experiment at the Relativistic Heavy Ion 
Collider (RHIC) for the systems of 
Au + Au and  Cu + Cu at \rootsnn{200} and at $62.4$~GeV, 
and p + p at \rootsnn{200}.
BRAHMS explores reaction dynamics 
and properties of the hot and high energy density matter produced 
in ultra-relativistic heavy-ion collisions versus its longitudinal expansion. 
Overall charged particle production, particle spectra over a large rapidity 
interval and \pt range are presented. Nuclear modification factors for 
Au~+~Au and Cu~+~Cu collisions are discussed. 
The observed number of charged particles produced per unit of 
rapidity in the central rapidity region indicates that a high energy density 
systems are produced at the initial stage of the Au + Au reaction. 
Analysis of anti-particle to particle ratios 
as a function of rapidity and collision energy
reveal that particle populations at the chemical freeze-out stage for 
heavy-ion reactions at and above
SPS energies are controlled by the baryon chemical potential.  
From the particle spectra we 
deduce significant radial expansion ($\beta \approx $ 0.75), as expected for systems
created with a large 
initial energy density.  
We also measure the elliptic flow parameter $v_2$ versus rapidity and \ptn. 
A weak dependence on rapidity
of the  \pt differential $v_{2}$ is observed.
We present rapidity dependent $p/\pi$ ratios within $0 < y < 3$ for Au~+~Au 
and Cu~+~Cu at \rootsnn{200}. 
The ratios are enhanced in nucleus-nucleus collisions as compared to p + p collisions. 
The particle ratios 
are discussed in terms of their system size and rapidity dependence.
We compare \Raa for Au + Au at \rootsnn{200} and  at $62.4$~GeV, and
for Au + Au and Cu + Cu at \rootsnn{62.4}. 
\Raa is found to increase with decreasing collision energy, decreasing system size, and
when going towards more peripheral collisions. However,
\Raa shows only a very weak dependence on rapidity (for $0 < y < 3.2$),
both for pions and protons. 
    
\end{abstract}

\section{INTRODUCTION}

Reactions between heavy nuclei provide a unique opportunity to produce and study
nuclear (hadronic) matter far from its ground state, at high densities 
and temperatures. From the onset of the formulation of the quark model and the first 
understanding of the nature of the binding and confining potential between
quarks about 30 years ago, it has been realized that at very high density and
temperature, hadronic matter may undergo a transition to a more primordial 
form of matter.
This proposed state of matter named the quark gluon 
plasma (QGP) \cite{Shuryak}, is 
characterized by a strongly reduced interaction among its 
constituents, quarks and gluons, such that the partons would exist in a nearly
free state~\cite{Collins}.  
Experimental attempts to create the QGP in the laboratory by colliding heavy nuclei 
have been carried out for more than 20 years.  
During this period, center of mass energies per pair of colliding nucleons
have risen steadily from the \rootsnn{1} domain of the Bevalac at LBNL, to 
energies of \rootsnn{5} at the AGS at BNL, and to \rootsnn{17} at the SPS 
at CERN. Although a number of signals suggesting the formation of a very dense
state of matter have been observed, no solid evidence for QGP formation was 
found at these lower energies. 

In mid-August 2001 systematic data collecting by the four RHIC 
experiments, namely BRAHMS \cite{brahms_www}, PHENIX \cite{phenix_www}, 
PHOBOS \cite{phobos_www} and STAR \cite{star_www}, began at the energy 
of \rootsnn{200}. The RHIC operations started a new era of studies 
of ultra-relativistic nucleus-nucleus collisions. 
One of the main results obtained up to the present at RHIC is the observation
of significant
elliptic flow in central Au + Au collisions.
The large flow signal, 
that is consistent with
the hydrodynamic evolution of a perfect fluid 
\cite{star_flow_papers,phenix_flow_paper,phobos_flow_paper,hydro1}, 
indicates a strongly interacting QGP, contrary to initial expectations.
Also observed is strong jet 
suppression \cite{br_PRL91}, which is predicted within the 
Quantum Chromo-Dynamics theory (QCD) as a consequence of
the creation of a dense colored medium \cite{khar_CGC}.
This last observation was supplemented by d~+~Au measurements
showing the absence of suppression, but rather a Cronin type enhancement 
at the central rapidity region, thus excluding an alternative interpretation 
of the suppression in terms of initial state parton saturation (CGC) effects.
Another measurement that strongly supported the scenario of jet quenching
was the disappereance of the away-side jet
\cite{star_Adler,star_Hardtke} in central Au~+~Au collisions,
whereas for d~+~Au and peripheral Au~+~Au near side and away side back-to-back
jet correlations have been measured.
Its large rapidity range and \pt coverage allows BRAHMS to study the properties 
of the produced medium as a function of its longitudinal expansion.
The measurement of rapidity evolution of the nuclear modification factor for d~+~Au
performed by BRAHMS shown that at more forward rapidities the hadronic yields are suppressed 
as compared to scaled p~+~p interactions \cite{br_evordau}. The suppression was even stronger 
for central collisions. Both observations can be quantitatively described
within the framework of the CGC \cite{khar_CGC,marian}.
 
\section{BRAHMS DETECTOR SETUP }   

The BRAHMS (Broad RAnge Hadron Magnetic Spectrometers) \cite{br_nim},
experimental setup consists of a set of global detectors and two spectrometer arms: 
a Mid-Rapidity Spectrometer (MRS) 
that operates in the polar angle range from 30 to 90 degrees, and a Forward Spectrometer (FS)
that operates in the range between 2.3 and 30 degrees.
Global detectors are used to measure the global features of the 
collision such as the overall particle multiplicity 
and collision centrality, position of the collision vertex and, more recently,  
information on 
the reaction plane orientation which can be used for azimuthal flow analysis. 
For the momentum measurements in the MRS we use two tracking devices and one dipole magnet.
Particle identification (PID) is done by Time-of-Flight (TOF) measurement and by using 
threshold Cherenkov detector C4.
In the FS two Time Projection Chambers (TPC) and three Drift Chambers (DC) deliver 
particle track segments to allow excellent momentum resolution using three dipole magnets. 
PID in the FS is provided by TOF measurements for low (TOF1) and medium (TOF2) particle momenta.
High momentum particles are identified using
a Ring Imaging Cherenkov detector (RICH). 
The BRAHMS PID ability is summarized in Table~\ref{tab_pid}. 
\begin{table}[htb]
\caption{Upper range of the momentum for 2$\sigma$ separation (in GeV/c)}
\label{tab_pid}
\newcommand{\m}{\hphantom{$-$}}
\newcommand{\cc}[1]{\multicolumn{3}{c}{#1}}
\newcommand{\ccc}[1]{\multicolumn{3}{c|}{#1}}
\renewcommand{\tabcolsep}{1pc} 
\renewcommand{\arraystretch}{1.2} 
\begin{tabular}{@{}lclclclclclcl}
\hline
{}       & \ccc{$0 < \eta <1.0$}   &  \cc{$1.5 < \eta < 4.0$} \\
\hline
{}       & TOFW    & TOFW2 & C4    & TOF1  &   TOF2   & RICH \\
\hline
$K/\pi$  & 2.0     & 2.5   & -     & 3.0   &   4.5    & 25.0 \\
$K/p$    & 3.5     & 4.0   & 9.0   & 5.5   &   7.5    & 35.0 \\
\hline
\end{tabular}
\end{table}

\section{OVERALL BULK CHARACTERISTICS}

\begin{figure}
   \centering
  \epsfig{file=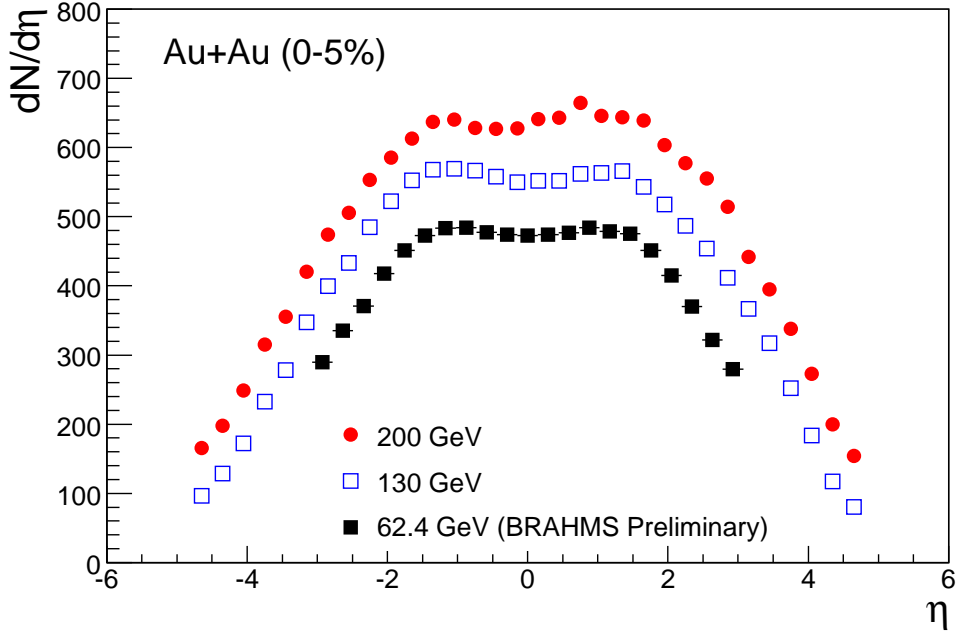,width=14.0cm}
  \vspace{-1.0cm}
  \caption{
    Distributions of  $dN_{ch}/d\eta$ measured by BRAHMS for $ 0-5\% $ central Au~+~Au reactions, 
    at \rootsnn{200} (solid circles), \rootsnn{130} (open squares) 
    and  \rootsnn{62.4} (solid squares). Only statistical errors (usually smaller than
    the symbol size) are shown.
  } 
  \label{dNdEta}  
\end{figure}

The multiplicity distribution of emitted particles is a fundamental observable in
ultra-relativistic collisions. It is sensitive to all stages of the reaction and
can address issues such as the role of hard scatterings between partons and the 
interaction of these partons in the high-density medium \cite{khar_multi,br_mult1,br_mult2}. 
Figure \ref{dNdEta} shows the measured pseudo-rapidity density of 
charged hadrons, \dndeta, over a wide range of $\eta$ for central $(0-5\%)$ Au~+~Au collisions
at \rootsnn{200}, \ene{130} and \ene{62.4}.
For central collisions at \rootsnn{200} we observe about 4500 
charged particles within the rapidity range covered by the detection system and 
\dndeta\nolinebreak$|_{\eta=0}$ = 625 $\pm$ 56. 
The latter value exceeds the particle production per participant 
pair observed in elementary p + ${\rm \bar{p}}$ collisions at the same 
energy by 40 - 50\% \cite{alner}. 
This means that nucleus-nucleus 
collisions at the considered energies are far from being the simple superposition of 
elementary nucleon-nucleon collisions.

The measurement of charged particle density $dN_{ch}/d\eta$ can be used to estimate
the so-called Bjorken energy density, $\varepsilon$ \cite{bjor}. The formula 
\begin{equation}
\label{eq1}
\varepsilon = \frac{3}{2} \times \frac{\langle E_{t}\rangle}{\pi R^{2}\tau_{o}}
                          \times \frac{dN_{ch}}{d\eta}
\end{equation}
provides the value of about 4 GeV/fm$^3$. To obtain this result we assumed that 
$\tau_{o}$~=~1~fm/c, $\langle E_{t}\rangle$ = 0.5 GeV and R = 6 fm.
The factor 3/2 is due to the assumption that the charged
particles carry out of the reaction zone only a fraction (2/3) of
the  total available energy. The more refined results obtained from  
identified particle abundances and particle spectra 
leads to somewhat larger values of $\varepsilon$, namely
5 GeV/fm$^3$ at \rootsnn{200}, 4.4~GeV/fm$^3$ at \rootsnn{130},
and  3.7~GeV/fm$^3$ at \rootsnn{62.4}, \cite{br_meson}. All of these values 
significantly exceed the predicted energy density $\varepsilon \approx 1$~GeV/fm$^3$ 
for the boundary between hadronic and partonic phases of nuclear matter \cite{karsch}.

\subsection{Hadrochemistry with BRAHMS data}

\begin{figure}[t]
  \centering
  \epsfig{file=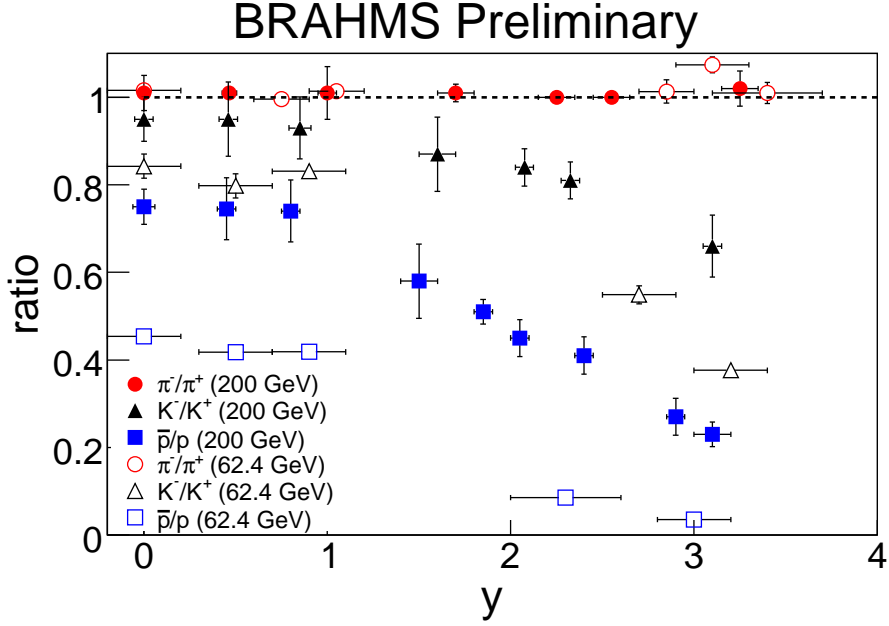,width=13.0cm}
  \vspace{-1.0cm}
  \caption{
   Ratios of anti-particles to particles (pions, kaons and protons) as a function of rapidity for
   \rootsnn{62.4}, \rootsnn{200}. Statistical and systematic errors are indicated. 
   }
   \label{ratiosVsY}  
\end{figure}

BRAHMS measures anti-particle to particle ratios for
pions, \pir, kaons, \kar, and protons, \pr, over a large
rapidity interval. The results, for Au~+~Au collision at \rootsnn{200} and \rootsnn{62.4} 
are presented in Figure \ref{ratiosVsY}.
Whereas \pionr stays constant and is equal to 1 over the covered 
rapidity range (0 $<$ y $<$ 3.2) for both energies,
the \kaonr and \protr ratios drop significantly with increasing rapidity.
For \rootsnn{200} the \kaonr and \protr ratios are equal, respectively to 0.95 and 0.76 
at $y \approx 0$, and reach values of 0.6 for \kaonr and 0.3 for \protr around rapidity 3.
Comparing \rootsnn{200} and \rootsnn{62.4} we observe a $11 \%$ and $40 \%$ decrease of the ratio 
at mid-rapidity and  a $43 \%$ and $93 \%$ decrease at $y \approx 3$ for 
kaons and protons, respectively.
The large change in the \protr ratio from  \rootsnn{200} to \rootsnn{62.4}  
is due to the fact that the rapidity $y \approx 3$ for the lower energy  corresponding
to the fragmentation region, whereas for \rootsnn{200}, $y \approx 3$ is located about 1 unit
of rapidity below the maximum in the net-baryon
distribution \cite{br_stopping}.
  
\begin{figure}[htb]
  \centering
  \epsfig{file=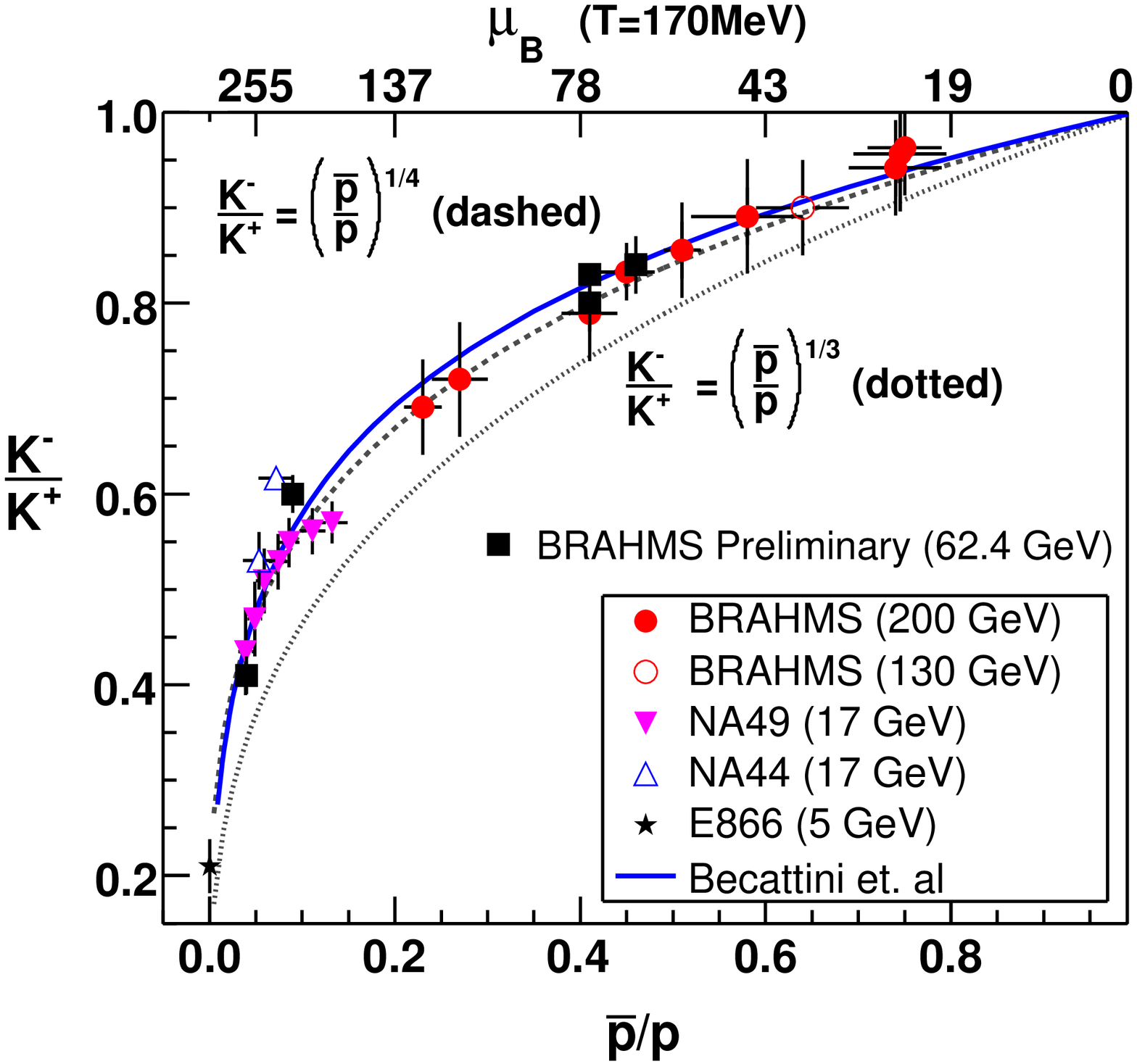,width=11.0cm}
  \vspace{-1.0cm}
  \caption{Correlation between \kar and \pr.
    The solid curve refer to statistical model calculation with a chemical freeze-out 
    temperature of 170 MeV.
  }
  \label{ratiosKaons}
\end{figure}

Figure \ref{ratiosKaons} shows the \kaonr ratio as a function of the corresponding
\protr for various rapidities. The presented results were obtained 
for central collisions at three RHIC 
collision energies. The AGS and SPS results
are plotted for comparison. There is a striking correlation between the 
RHIC/BRAHMS kaon and proton ratios over 3 units of rapidity.
%
%
It is worth noting that the BRAHMS forward rapidity data measured at \rootsnn{62.4} overlap
with the SPS points that were measured at much lower energy but at mid-rapidity.
The solid line in Figure 3 shows a fit with a statistical model to the
\rootsnn{200} results, only, assuming
that the temperature at the chemical freeze-out is 170 MeV \cite{br_ratios,becatt}.
It is seen that the data are very well described by the statistical model over a broad 
rapidity range with the baryon chemical potential changing from 27 MeV at mid-rapidity 
to 140 MeV at the most forward rapidities. 
Using simple statistical models at the quark level with chemical and thermal 
equilibrium, the ratios can be written 
\begin{equation}
\label{eq2}
\frac{N_{\bar{p}}}{N_p}= e^{-6\mu_{u,d}/T}, \; \; \; \; \; \; \; \; 
\frac{N_{K^-}}{N_{K^+}} = e^{-2(\mu_{u,d} - \mu_{s})/T},
\end{equation}
where $\mu$ and $T$ are the chemical potential and temperature, respectively.
Substituting $\mu_s = 0$ into eqs. (\ref{eq2}), one gets 
\kaonr = [\pr]$^{1/3}$. This relation, represented by the dotted line
on Figure \ref{ratiosKaons}, does not reproduce the observed correlation.
The data are, however, well fitted by the function  
\kaonr = [\pr]$^{1/4}$ (dashed line) which can be 
derived from eqs. (\ref{eq2}) assuming $\mu_s = 1/4 \mu_{u,d}$.

Recently, STAR and NA49 have measured mid-rapidity ratios
$\bar{\Lambda}/\Lambda$, $\bar{\Xi}/\Xi$ and $\bar{\Omega}/\Omega$ 
versus \protr for a set of energies from \rootsnn{10} up to  \rootsnn{200}. 
These preliminary results can also be well described  within a statistical 
model of chemical and thermal equilibrium at the quark level and confirm
the strong correlation between $\mu_s$ and $\mu_{u,d}$ derived from
BRAHMS data.
%
 

\subsection{Radial flow}

\begin{figure}[htb]
\begin{minipage}[t]{80mm}
\centering
\epsfig{file=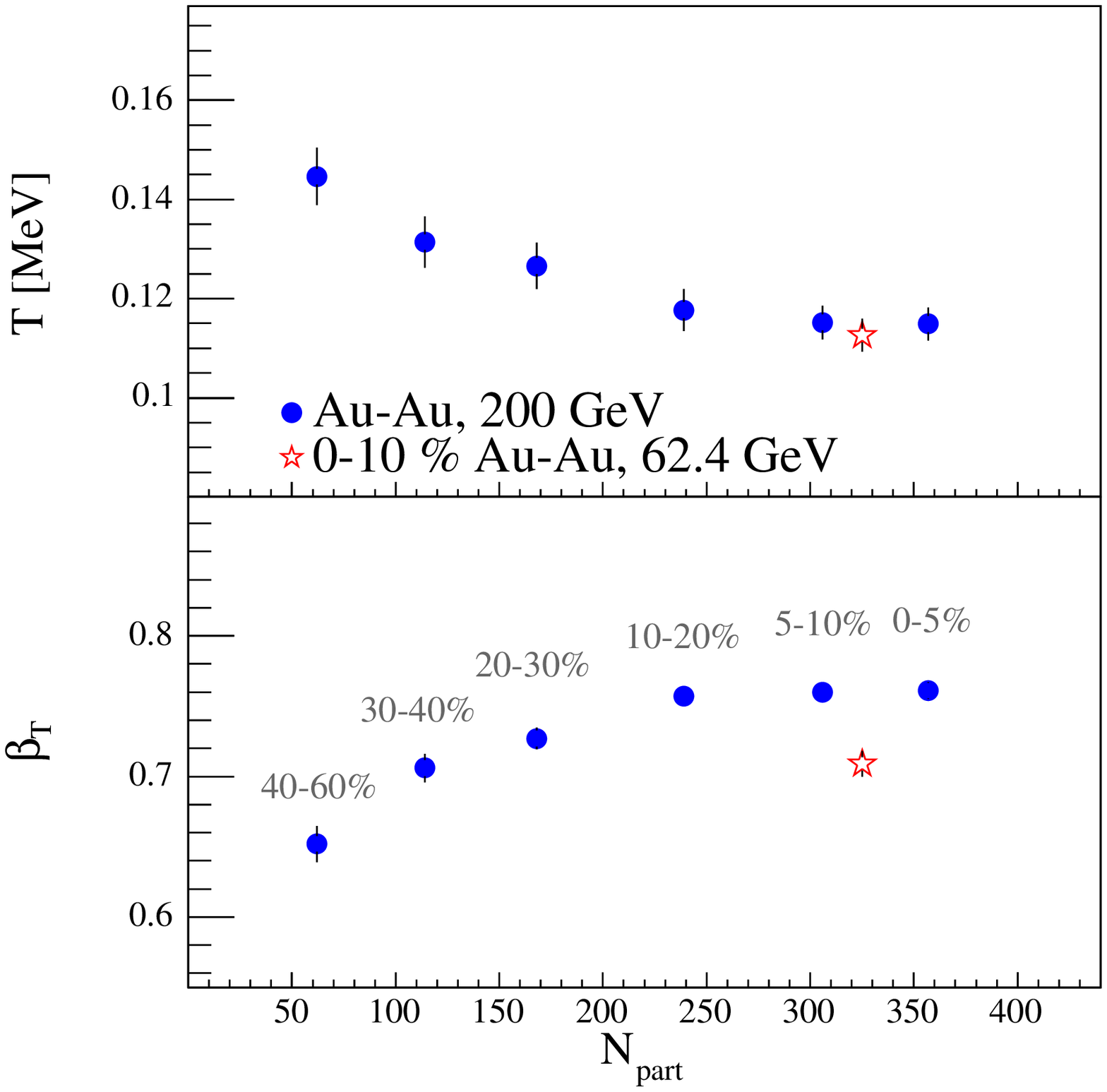,width=8.3cm}
 \vspace{-1.0cm}
\caption{ Kinetic freeze-out temperature and transverse flow velocity at mid-rapidity as a function 
    of centrality for Au~+~Au at \rootsnn{200} (dots). For comparison
    we show the result for $0-10\%$ central Au~+~Au collisions at \rootsnn{62.4} (stars).}
\label{flowNpart} 
\end{minipage}
\hspace{\fill}
\begin{minipage}[t]{70mm}
\centering
\epsfig{file=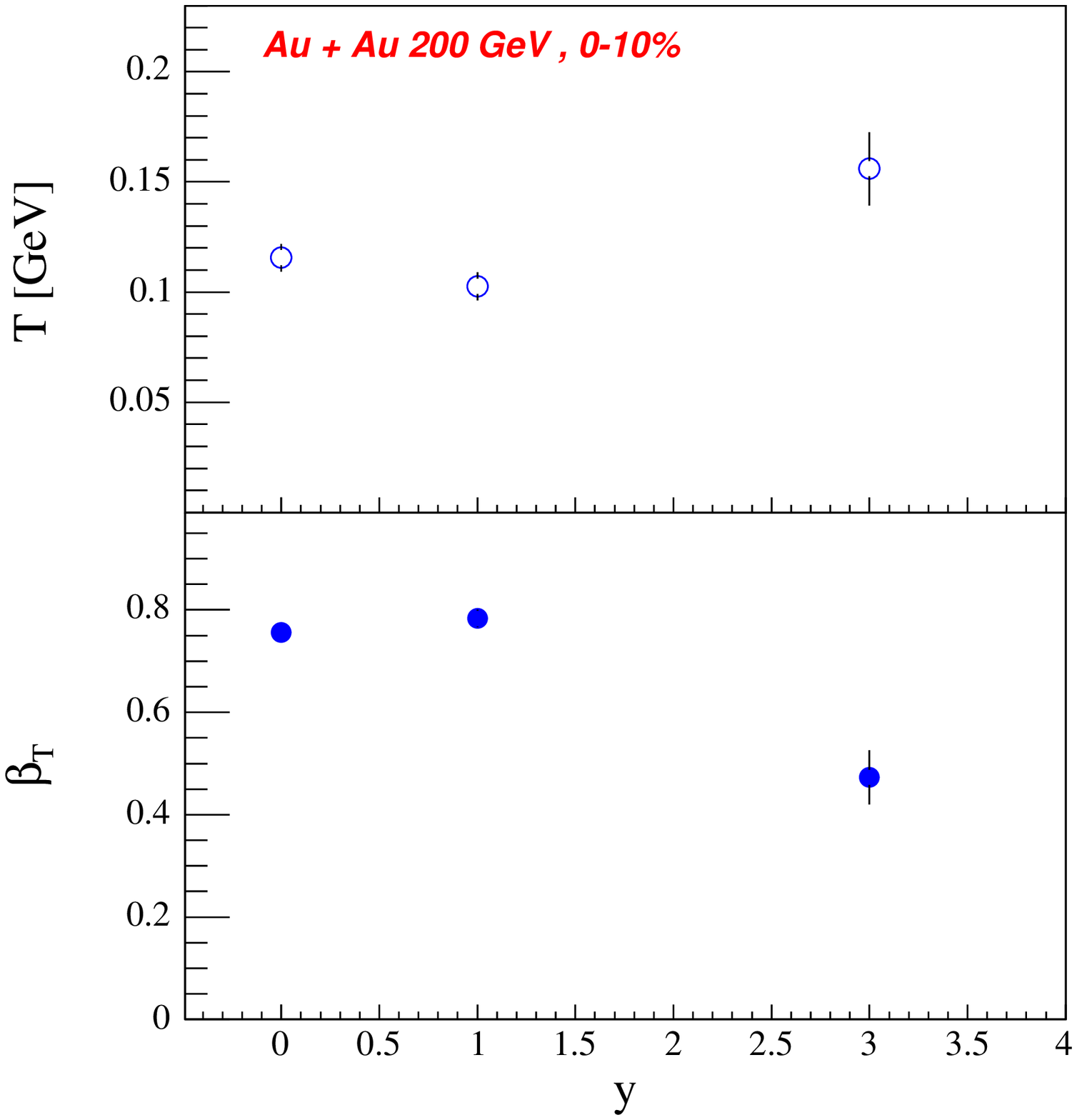,width=7.8cm}
 \vspace{-1.0cm}
 \caption{ Kinetic freeze-out temperature and transverse flow velocity for central Au + Au 
   collisions  at \rootsnn{200} as a function of rapidity.
 }
 \label{flow_y}
\end{minipage}
\end{figure}

The properties of matter in the latest stage of the collision when the
interactions between particles cease (kinetic freeze-out) 
can be studied from the shapes of emitted particle spectra.
These shapes depend in general on the temperature of the emitting source and on the
collective flow. For central collisions where one should not
expect any azimuthal dependence, only the so-called transverse flow is 
important \cite{hydro1}.
In the blast-wave approach \cite{blast1} the spectrum is parametrized by a function 
which depends on the freeze-out temperature, $T_{fo}$, and on 
the transverse expansion velocity, $\beta_{T}$.
Figure \ref{flowNpart} shows results obtained from the analysis 
of the Au + Au reactions at \rootsnn{200}.
The blast-wave model was used to fit, simultaneously, the $\pi^+$, $\pi^-$,
K$^+$, K$^-$, protons and anti-protons spectra. 
The results are plotted versus the number of participants, $N_{part}$.

One can see that the kinetic freeze-out temperature 
decreases with centrality from about 140 MeV for the $40-50\%$ centrality bin
to about 120 MeV for the most central collisions. 
The latter value is lower than the temperature of chemical freeze-out 
indicating that, as expected, the freeze-out of particle ratios occurs earlier 
than the kinetic freeze-out.
The reversed trend is seen
for the expansion velocity which is equal to about 0.65c and 0.75c for
$40-50\%$ and $0-5\%$ collision centrality, respectively.
For comparison we show also the result for the central Au + Au collisions at \rootsnn{62.4}.
Although for the same value of $N_{part}$ the $T_{fo}$ value for both energies is very similar,
a reduction by about $20 \%$ in the expansion velocity is observed. 
 
Figure \ref{flow_y} shows the dependence of $T_{fo}$ and $\beta_{T}$ 
as a function of rapidity. The observed variations of $T_{fo}$ and $\beta_T$ on 
$N_{part}$ and rapidity are consistent with the
hydrodynamic description in which the radial flow is the result of outwards 
gradients of pressure that exist in the
expanding matter during the whole evolution. 
Thus the speed of expansion should increase with
the density of the initially created system.    
The transverse expansion velocity is larger than that observed at SPS energies, which
is consistent with a large initial density of the system created at
RHIC.   

\subsection{Elliptic Flow}

A powerful tool for studying the dynamics that drive the initial evolution of
systems created in heavy ion reactions is the analysis of the azimuthal
distribution of the emitted particles relative to the reaction plane \cite{hydro1}.
The triple differential distribution of emitted particles can be factorized
as follow
\begin{equation}
\frac{dN}{dydp_Td\phi} = \frac{dN}{dydp_T} \frac{1}{2\pi} \times
(1 + 2v_1(y,p_T)cos(\phi-\phi_r) + 2v_2(y,p_T)cos2(\phi-\phi_r) + ....),
\label{eq_v2}
\end{equation}
where $\phi$ and $\phi_r$ denote the azimuthal angles of the particle
and of the reaction plane, respectively. 
The first factor depends only on $y$ and $p_T$ and the
second factor represents the expansion of the azimuthal dependence
into a Fourier series. The coefficients of 
the first ($v_1$) and the second ($v_2$) harmonics are called
direct and elliptic flow parameters, respectively, and in general
they are functions of rapidity and transverse momentum. 
Calculations based on hydrodynamic models \cite{hydro1}
show that elliptic flow is substantially generated only during the 
highest density phase, before the initial spatial anisotropy of the
created medium disappears. Thus $v_2$ is sensitive to the very
early phase in the system evolution and relatively insensitive
to the late stage characterized by the dissipative expansion of the hadronic gas.

Elliptic flow has been extensively measured by the STAR, PHENIX and PHOBOS
experiments \cite{star_flow_papers,phenix_flow_paper,phobos_flow_paper}.
Generally, $v_2$ is an increasing function of \pt up to
$1.5$~GeV/c, at which point it saturates. Up to roughly 1.5 GeV/c in \ptn,
hydrodynamic calculations show good agreement with experimental data for the $v_2$
dependence on \pt and centrality.

\begin{figure}
  \centering
  \epsfig{file=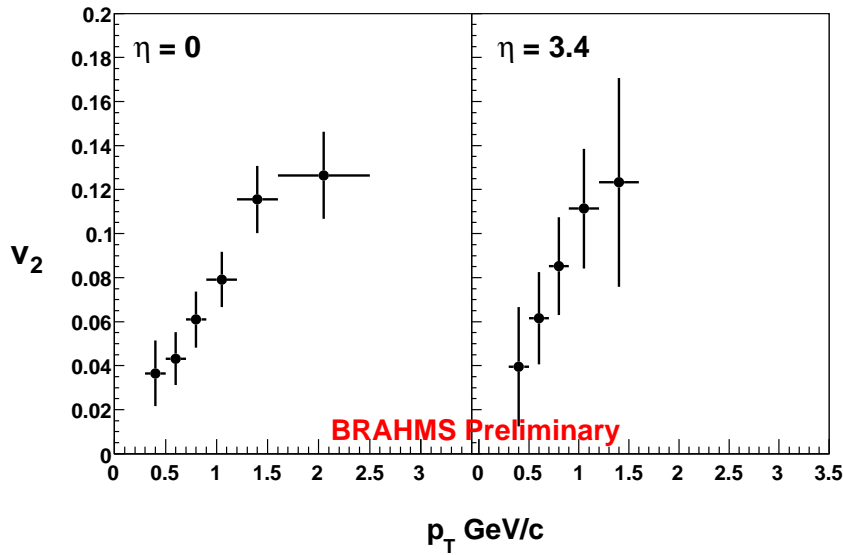,width=12.0cm}
  \vspace{-1.0cm}
  \caption{
    Pion elliptic flow parameter \pt dependence measured for Au + Au at \rootsnn{200}
    at $\eta = 0$ and at $\eta = 3.4$.}
  \label{v2_ypt}  
\end{figure}

BRAHMS has measured the \pt dependence of $v_2$ at a number of rapidities \cite{qm05_hiro}, 
and the results for pions at $\eta = 0$ and at $\eta = 3.4$ are shown in Fig. \ref{v2_ypt}.
The elliptic flow parameter, $v_2$, for identified pions is an increasing function 
of \pt at both rapidities. 
The dependence on pseudo-rapidity is very small.
These results are very similar to those obtained for charged hadrons \cite{qm05_hiro}.

\section{BARYON TO MESON RATIOS}

\begin{figure}[htb]
\begin{minipage}[t]{65mm}
 \epsfig{file=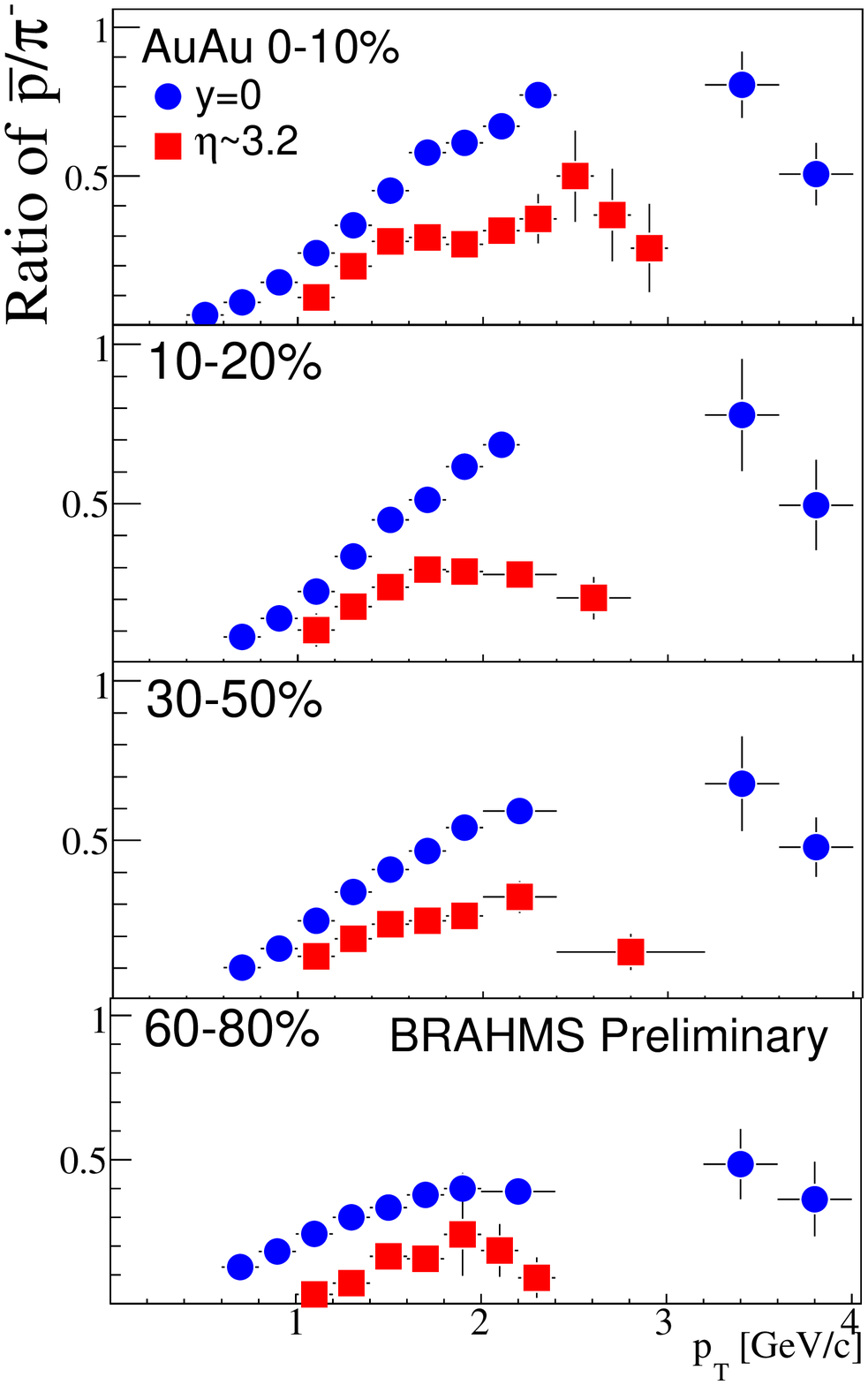,width=6.5cm}
  \vspace{-1.0cm}
  \caption{The $\bar{p}$ to $\pi^-$ ratios at mid-rapidity (circles) and at 
    pseudo-rapidity $\eta \approx 3.2$ (squares) for different centralities of Au + Au collisions at 
    \rootsnn{200}.
    }
  \label{btom1}  
\end{minipage}
\hspace{\fill}
\begin{minipage}[t]{85mm}
 \epsfig{file=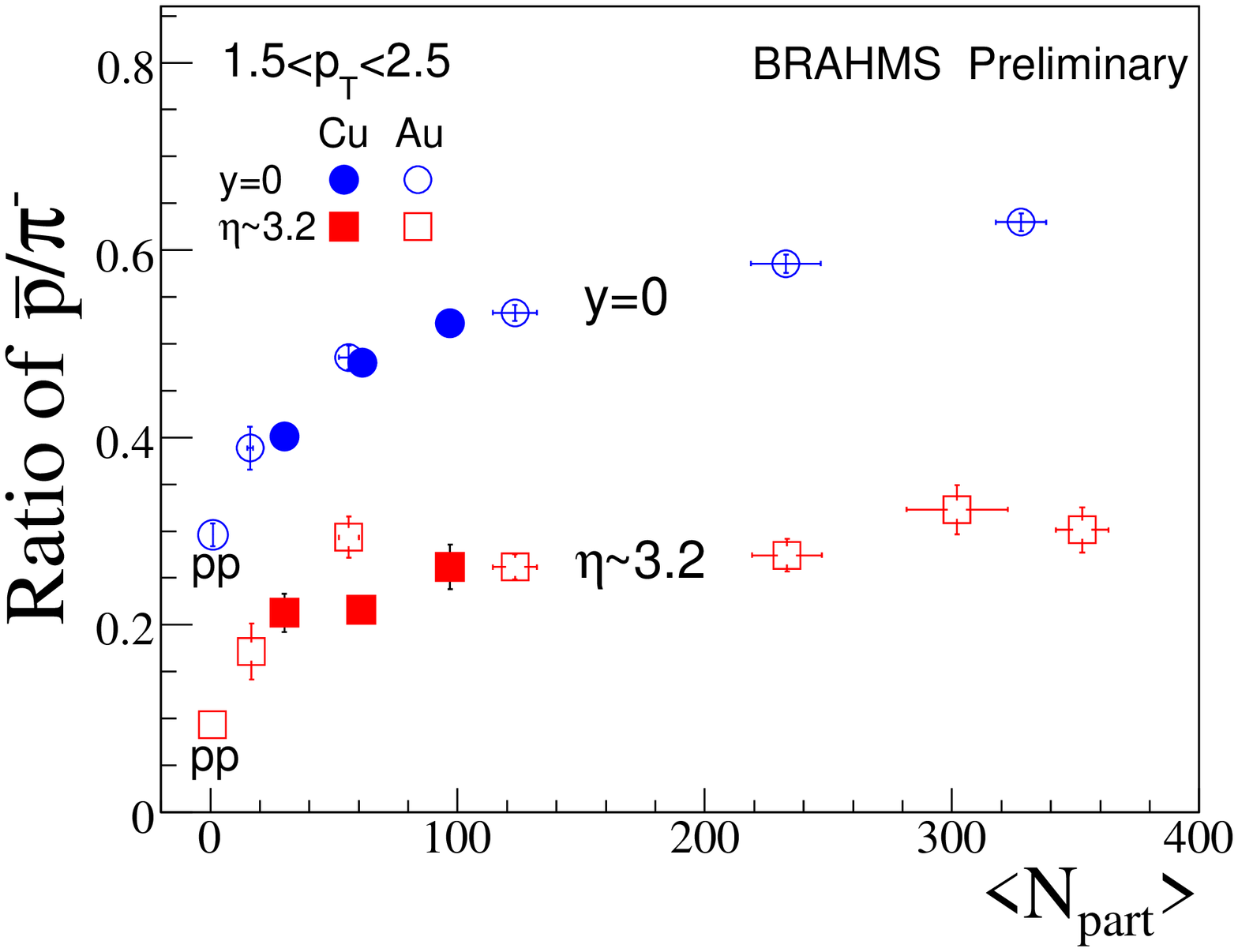,width=8.5cm}
  \vspace{-1.0cm}
  \caption{The averaged $\bar{p}/\pi^-$ versus $\langle N_{part} \rangle$ for Au + Au (open
    symbols) and Cu + Cu (solid symbols) at \rootsnn{200}, for $y \approx 0$ (circles) and
    and for $\eta \approx 3.2$ (squares).
    }
  \label{btom2}  
\end{minipage}
\end{figure}

\begin{figure}
  \centering
  \epsfig{file=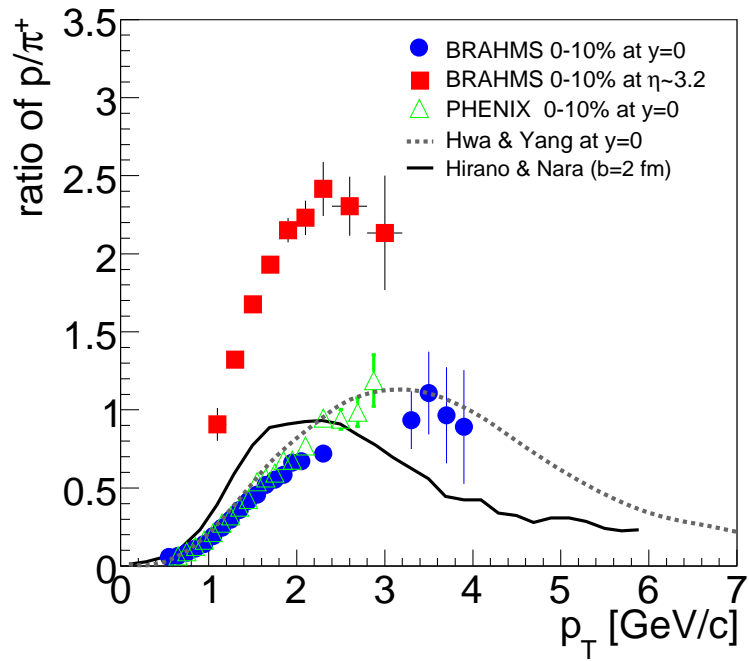,width=10.5cm}
  \vspace{-1.0cm}
  \caption{The $p/\pi^+$ ratio versus \pt for Au + Au collisions  at \rootsnn{200}
    measured at mid-rapidity by BRAHMS (circles) and by PHENIX (triangles).
    Squares represent BRAHMS results at $\eta \approx 3.2$.
    The plotted lines show model calculations (see text). 
  }
  \label{btom3}  
\end{figure}

With its excellent particle identification capabilities, BRAHMS can study
the \pt and $y$ dependence of hadron production.
Preliminary results \cite{br_Claus,br_Yin} indicate that for Au~+~Au reactions in the intermediate
\pt region the proton to meson ratio is significantly higher that one would
expect from the parton fragmentation in vacuum process. Several theoretical
models have been proposed to explain the observed enhancement.
These range from models that explore the 
partonic interactions of quark hadronization and quark coalescence \cite{eunjoo_Fries,eunjoo_Greco,eunjoo_Hwa}
to models that incorporate novel baryon dynamics \cite{eunJoo_Vitev,VTPop} .
The recent experimental data obtained for
the p~+~p, Au~+~Au, and Cu~+~Cu colliding systems are expected to result in a 
better understanding of the underlying physics. 
Fig. \ref{btom1} shows a recent investigation by BRAHMS \cite{qm05_eunjoo} of the
$\bar{p}$ to $\pi^-$ ratios at mid-rapidity (circles) and at pseudo-rapidity $\eta \approx 3.2$ (squares)
for Au + Au collisions at \rootsnn{200} for the different centrality classes indicated on the plot. 
The data reveal a smooth increase of $\bar{p}/\pi^-$ ratios from peripheral to central collisions, 
however, the centrality dependence is stronger at mid-rapidity than at forward rapidity.
The maxima in the $\bar{p}/\pi^-$ ratios are lower at forward rapidity as compared to mid-rapidity.   
Figure~\ref{btom2} shows the $\bar{p}/\pi^-$ centrality dependence for Au~+~Au (open
symbols) and Cu~+~Cu (solid symbols) at \rootsnn{200}. The data for $y = 0$ and $\eta \approx 3.2$
are represented by circles and squares, respectively. 
One can see the strong increase of the 
$\bar{p}/\pi^-$ ratios as a function of \Npart in the range  $0 < N_{part} < 60$. 
For $N_{part} > 60$ the dependence starts to saturate 
and the ratios reach values of about 0.6 and about 0.25 for central collisions at mid- and 
forward rapidities, respectively.  
For peripheral Au~+~Au collisions the data approach the p~+~p results.
It is important to note that the Au + Au and Cu + Cu ratios are consistent with each other
when plotted versus $\langle N_{part} \rangle$, indicating that the $\bar{p}/\pi^-$ ratio  
is controlled by the initial size of the created systems.
Fig. \ref{btom3} shows the comparison of BRAHMS and PHENIX data for the ratio of protons to positive
pions measured at $y=0$. 
The parton coalescence \cite{eunjoo_Greco} and recombination \cite{eunjoo_Hwa} 
models describe the observed ratios well at mid-rapidity, as shown in the figure
by the dashed curve.
The three-dimensional hydrodynamic model \cite{eunjoo_Hirano_Nara} (solid curve)
cannot reproduce the observed shape of the $p/\pi^+$ ratio, however it reproduces
the overall level of enhancement. 
The current hydrodynamic calculations do not show a rapidity dependence which is
contrary to the experimental observation (see Fig. \ref{btom3}).
This is presumably because the hydrodynamic models 
assume constant baryon chemical potential versus rapidity.

\section{HIGH \pt SUPPRESSION}
Particles with high \pt (above 2 GeV/c) are primarily produced in hard 
scattering processes early in the collision. In high energy nucleon-nucleon 
reactions hard scattered partons fragment into jets of hadrons.
However, in nucleus-nucleus collisions hard scattered partons might 
travel in the medium. It was predicted that if the medium is a QGP, the partons
will lose a large fraction of their energy by induced gluon
radiation, effectively suppressing the jet production \cite{wang}.
Experimentally this phenomenon, known as jet quenching, will be observed 
as a depletion of the high \pt region in hadron spectra. 

The measure most commonly used to study the medium effects is called 
the nuclear modification factor, \Rnn. It is defined as the ratio of the 
particle yield produced in nucleus-nucleus collision, 
scaled with the number of binary collisions (\Nncoll), and the particle yield 
produced in elementary nucleon-nucleon collisions:
\begin{equation}
\label{eq5}
 R_{AA} = \frac{Yield(AA)}{N_{coll} \times Yield(NN)}.
\end{equation}
At high \pt the particles are predominantly 
produced by hard scatterings and in the 
absence of nuclear effects (when the nucleus-nucleus collision reduces to the 
superposition of elementary collisions) we expect \Raa to be 1. At 
low $p_T$, where the production rate scales rather with \Nnpart, \Raa 
should converge to N$_{part}$/\Ncoll which is roughly 1/3 
for central Au~+~Au collisions at the top RHIC energy.  
\Raa $< 1$ at high \pt will indicate the suppression which, 
as has been discussed, 
can be attributed to the jet quenching phenomenon. At SPS there is no suppression, 
in fact it is well known that there is enhancement for \pt~$>$~2~GeV/c.  
This so-called Cronin effect is attributed to initial multiple scattering 
of reacting partons.

Another variable used to quantify nuclear effects, one that does not depend on the
elementary reference spectra, is \Rncp, defined as the ratio of
\Raa for central nucleus-nucleus collisions to \Raa for peripheral collisions.
The idea of using \Rcp is based on the expectation that any nuclear modification in peripheral
collisions will be insignificant. We will show that the latter
statement is not true for $R_{CP} = R_{AuAu}(0-10\%)/R_{AuAu}(40-60\%)$. 

\subsection{\Raa evolution on collision centrality and collision energy for Au~+~Au and Cu~+~Cu systems}

\begin{figure}
  \centering
  \epsfig{file=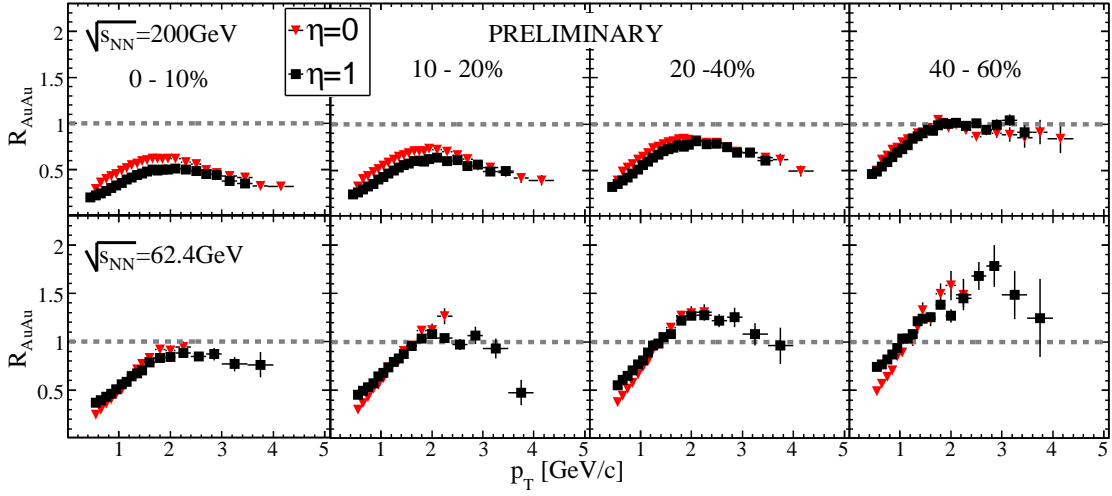,width=15.0cm}
 \vspace{-1.0cm}
  \caption{\Raa measured at $\eta = 0$ and  $\eta = 1$ for Au + Au at 
    \rootsnn{200} (upper row) and at \rootsnn{62.4} (bottom row), for different 
    centrality bins indicated on the plot 
    (p + p reference for  \rootsnn{62.4} is based on ISR collider data \cite{Alper}).
  }
  \label{auau200vs62}  
\end{figure}

The large set of data collected during the RHIC run 4 has allowed us to carry
out the study of the \Raa evolution on collision centrality and collision
energy for two colliding systems, namely Au~+~Au and Cu~+~Cu \cite{qm05_truls}.
Figure \ref{auau200vs62} shows \Raa measured at $\eta = 0$ and  $\eta = 1$ 
for charged hadrons produced in Au + Au reactions at \rootsnn{200} 
(upper row) and at \rootsnn{62.4} (bottom row), for different 
collision centralities indicated on the plot.
For the most central reactions \Raa shows suppression for both energies,
however, the suppression is significantly stronger at the higher energy. 
We observe a smooth increase of \Raa towards less central collisions,
for \rootsnn{200}, resulting in approximate scaling with \Ncoll 
for $p_T > 2$ GeV/c for the $40-50 \%$ centrality bin. 
However, at \rootsnn{62.4}, \Raa $\approx 1$ for more central collisions, and
a Cronin peak is clearly visible already for the $20-40 \%$ centrality class, where \Raa reaches value 
of about 1.3 in the \pt range between $2.0$ and $3.0$ GeV/c.
These observations at \rootsnn{62.4} are qualitatively consistent with a picture 
in which there are two competing
mechanisms that influence the nuclear modification in the intermediate and high \pt range,
namely: jet quenching that dominates at central collisions and
Cronin type enhancement ($k_{T}$ broadening or/and quark recombination) 
that prevails for the more peripheral collisions.   
     
\begin{figure}
  \centering
  \epsfig{file=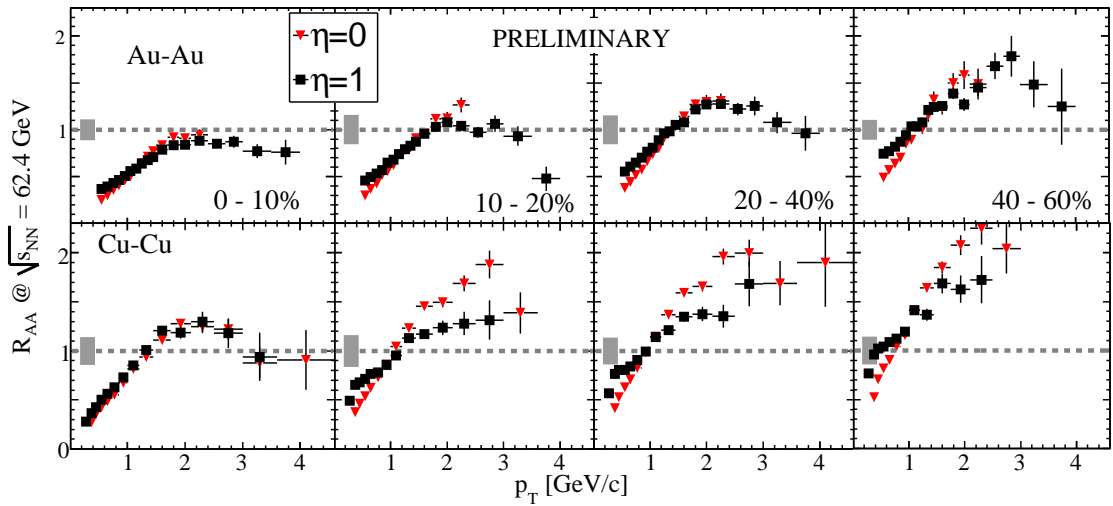,width=15.0cm}
  \vspace{-1.0cm}
  \caption{\Raa measured at mid-rapidity for Au + Au (upper row) and for Cu + Cu
    (bottom row) at \rootsnn{62.4} for different centrality ranges indicated on the plot.
  }
  \label{auauvscucu62}  
\end{figure}

Figure \ref{auauvscucu62} presents similar comparison as shown 
on Fig. \ref{auau200vs62} but this time we compare two different systems, namely 
Au~+~Au and Cu~+~Cu colliding at the same energy (\rootsnn{62.4}).
For Cu + Cu at \rootsnn{62.4} the same trend of increasing \Raa with decreasing 
collision centrality is seen. However, now the Cronin type
enhancement is present already for the most central collisions. 

Summarizing the whole set of observation we conclude that
the level of suppression of the inclusive hadron spectra produced in nucleus-nucleus collisions 
at RHIC energies in the \pt range above $2$ GeV/c increases with increasing collision energy,
collision centrality and with the size of the colliding nuclei. The dependency on
the last two variables can be replaced by only one dependency on \Npart \cite{qm05_truls}.
  

\subsection{\Raa for identified hadrons at forward rapidity for Au + Au at \rootsnn{200}}

The BRAHMS collaboration has already noted that the nuclear modification 
factor in central  Au + Au at \rootsnn{200} at  $\eta = 2.2$ is comparable to that 
measured at mid-rapidity \cite{br_PRL91} for the same system and energy. 
However, the data did not clearly identify the mechanism responsible for the observed effect.
In this section we present more exclusive analysis of nuclear effects at forward rapidity
for identified pions, kaons and protons.

\begin{figure}
  \centering
  \epsfig{file=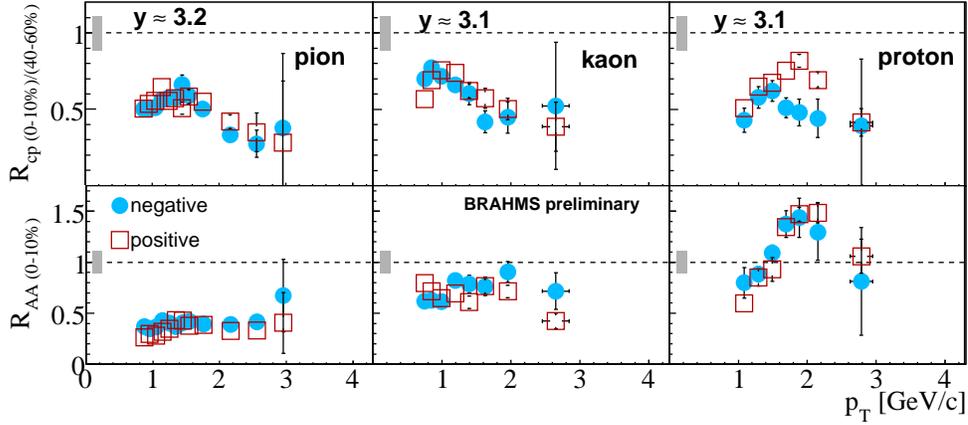,width=13.0cm}
  \vspace{-1.0cm}
  \caption{BRAHMS \Raa (upper row) and \Rcp (bottom row) for pions, kaons and protons, 
    measured $y \approx 3.2$ in Au + Au \rootsnn{200}. The p + p reference was also 
    measured by BRAHMS (see \cite{qm05_radek}).  
  }
  \label{AuAuRaaRcp}  
\end{figure}

Figure~\ref{AuAuRaaRcp} shows \Raa for identified hadrons at rapidity $y \approx 3.2$ for $0-10\%$ 
central Au~+~Au events at \rootsnn{200}. The shaded band around unity indicates systematic
error associated with the uncertainty in the number of binary collisions. 
Both \Raa and \Rcp
show suppression for pions (left panel) and for kaons (middle panel), however for
protons (right panel) \Rcp shows suppression whereas \Raa reveals a Cronin type enhancement
with the peak at $p_T \approx 2$ GeV/c. 
The difference between \Raa and \Rcp for protons is striking, indicating significant enhancement of protons
(in respect to p + p) for the $40-60 \%$ collision centrality used in the definition of \Rncp.
The same misleading behavior of \Rcp is seen for charged hadrons when
comparing evolution of \Raa and \Rcp on $\eta$ for Au + Au at 
\rootsnn{200} and \rootsnn{62.4} \cite{qm05_truls}.
  
\begin{figure}[htb]
  \centering
  \epsfig{file=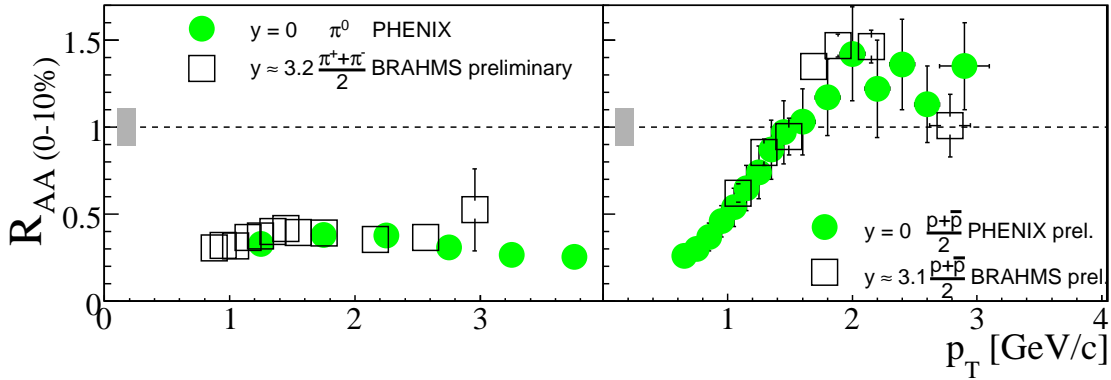,width=15.0cm}
  \vspace{-1.0cm}
  \caption{
    Comparison of \Raa measured for central Au + Au collisions at \rootsnn{200}, 
    at mid-rapidity and $y \approx 3$ for pions (left panel) and protons (right panel).
  }
  \label{PhenixRaa}
\end{figure}

Figure~\ref{PhenixRaa} shows the nuclear modification factors calculated for
$(\pi^{+} + \pi^{-})/2$ (left panel) and $(p + \bar{p})/2$ (right panel), 
respectively, at $y \approx 3.2$, for central Au + Au reaction. 
For the comparison we also show the \Raa values measured by the PHENIX Collaboration at mid-rapidity. 
The \Raa measured for pions shows strong suppression (by factor of about 3 for $2 < p_T < 3$ GeV/c),
both at mid- and at forward rapidity. The consistency between mid- and forward rapidity is seen also
for protons, but in this case, \Raa reveals a Cronin peak around $p_T = 2$ GeV/c.         
The similarity between \Raa at mid- and forward rapidity observed simultaneously for
pions and protons suggests the same mechanisms is responsible for the nuclear modifications
within the studied rapidity interval.
  
\begin{figure}[htb]
  \centering
  \epsfig{file=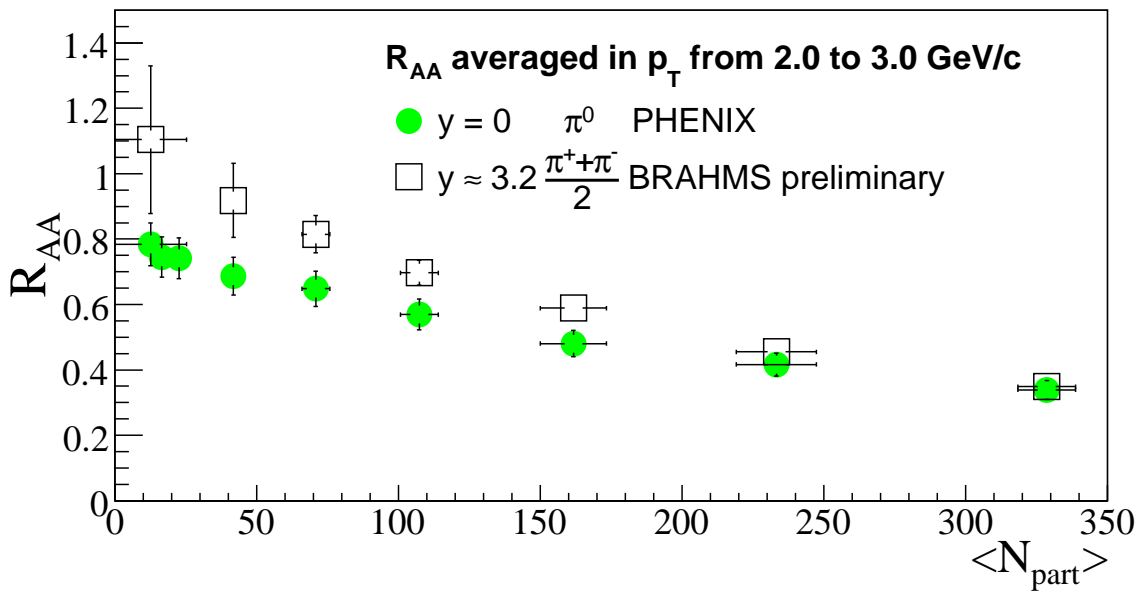,width=15.0cm}
  \vspace{-1.0cm}
  \caption{Averaged \Raa in the range $2.0 < p_T < 3.0$ GeV/c at mid-rapidity (PHENIX) and at forward
    rapidity as a function of collision centrality, expressed
    by the number of participants.
  }
  \label{AuAuRaaNpart}     
\end{figure}

It has been predicted \cite{gyulassy_quenching} that the magnitude of jet quenching 
should depend on both the size and the density of the created absorbing medium, thus it is interesting
to study the dependence of \Raa on centrality.
In Figure~\ref{AuAuRaaNpart} we plot the averaged \Raa measured for pions 
versus $\langle N_{part} \rangle$ for mid-rapidity (solid circles) and for 
forward rapidity (open squares). 
The averaging was performed in the \pt range from $2$ GeV/c to $3$ GeV/c.   
Once again, the mid- and forward rapidity pion suppression
for the most central Au + Au reaction are found of the same strength. 
However, the \Raa measured at forward 
rapidity shows significantly stronger rise towards peripheral collisions as 
compared to \Raa at mid-rapidity, differing on the level 
of $35 \%$ for $\langle N_{part} \rangle \approx 100$. The trend is consistent with the model of parton 
energy loss in a strongly absorbing medium \cite{deinese,drees}.
In this picture, for $y = 0$, the jet emission is dominated by the emission from the surface 
which quenches the dependence of \Raa on the system size. 
On the other hand, for $y \approx 3$, the transition from surface to volume emission 
can occur, which leads to a stronger dependence on \Nnpart.

\section{SUMMARY}

The results from BRAHMS and the other RHIC experiments clearly
show that studies of high energy nucleus-nucleus collisions have moved to a 
qualitatively new physics domain. The collisions are characterized by a high degree of reaction
transparency leading to the formation of a near-baryon-free central region
with approximate balance between matter and antimatter.
From the measurement of charged particle multiplicities in this region
lower limits for the energy density at $\tau_{o}$~=~1~fm/c have been determined 
as 5~GeV/fm$^3$ and 3.7~GeV/fm$^3$, for
central Au + Au reactions at \rootsnn{200} and \rootsnn{62.4}, respectively.
Therefore the conditions necessary for the formation of a deconfined system 
appear to be well fulfilled at RHIC energies.
Analysis within the statistical model
of the relative abundances of $K^{-}$, $K^{+}$, $p$ and $\bar{p}$ suggests 
equilibrium at a chemical freeze-out temperature of 170 MeV, with a noticeably
strong correlation between the strange quark and baryon chemical potentials.
Analysis of particle spectra within the blast-wave model indicates a kinetic 
freeze-out temperature on the order of 120 MeV and a large transverse expansion 
velocity, which is consistent with the high initial energy density.

The measurement of the elliptic flow parameter $v_2$ versus rapidity and $p_T$ 
shows weak dependence 
of the \pt differential $v_{2}$ on rapidity.
The $p/\pi$ ratios were measured within $0 < \eta < 3$ for Au~+~Au and Cu~+~Cu at
\rootsnn{200} and \rootsnn{62.4}. 
The obtained results reveal strong enhancement of baryon to meson
ratios in nucleus-nucleus collisions as compare to p + p collisions. 
Models that incorporate an interplay between
soft and hard processes can describe the data at mid-rapidity. 
We compared \Raa for Au~+~Au at \rootsnn{200} and \rootsnn{62.4}, and
for Au~+~Au and Cu~+~Cu at \rootsnn{200}. 
The general observed trend is that
\Raa increases with: decreasing collision energy, decreasing system size,  and
when going towards the more peripheral collisions. 
For Au~+~Au central collisions
at \rootsnn{200}, \Raa shows very weak dependence on rapidity (in $0 < y < 3.2$ interval),
both for pions and protons.

\end{document}